\documentclass[11pt]{article}

\usepackage[final]{acl}

\usepackage{times}
\usepackage{latexsym}

\usepackage[T1]{fontenc}

\usepackage[utf8]{inputenc}

\usepackage{microtype}

\usepackage{inconsolata}

\usepackage{graphicx}
\usepackage{placeins}
\usepackage{booktabs} 
\usepackage{hyperref}
\usepackage{multirow}
\usepackage{makecell}
\usepackage{tabularx}
\usepackage[table]{xcolor}
\usepackage{xcolor}
\definecolor{codegray}{gray}{0.93}
\usepackage{enumitem}

\usepackage{array}
\usepackage{siunitx}
\usepackage{amsmath}
\usepackage{amssymb}
\usepackage{mathtools}
\usepackage{amsthm}
\usepackage{arydshln}

\usepackage[utf8]{inputenc}
\usepackage{tcolorbox}
\usepackage{natbib}
\definecolor{nmgray}{HTML}{666666}

\newcommand{\graynote}[1]{\textcolor{black!40}{\emph{#1}}}

\newcommand{\providericon}[2][0.9em]{%
  \raisebox{-0.15ex}{\includegraphics[width=#1]{#2}}%
}

\newcommand{\anthropic}{\providericon[0.85em]{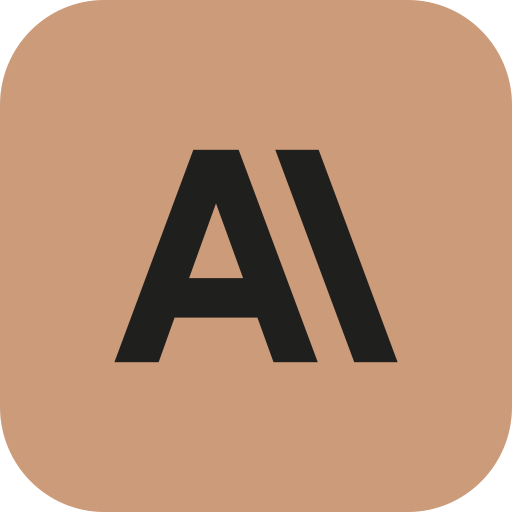}}
\newcommand{\openai}{\providericon[0.85em]{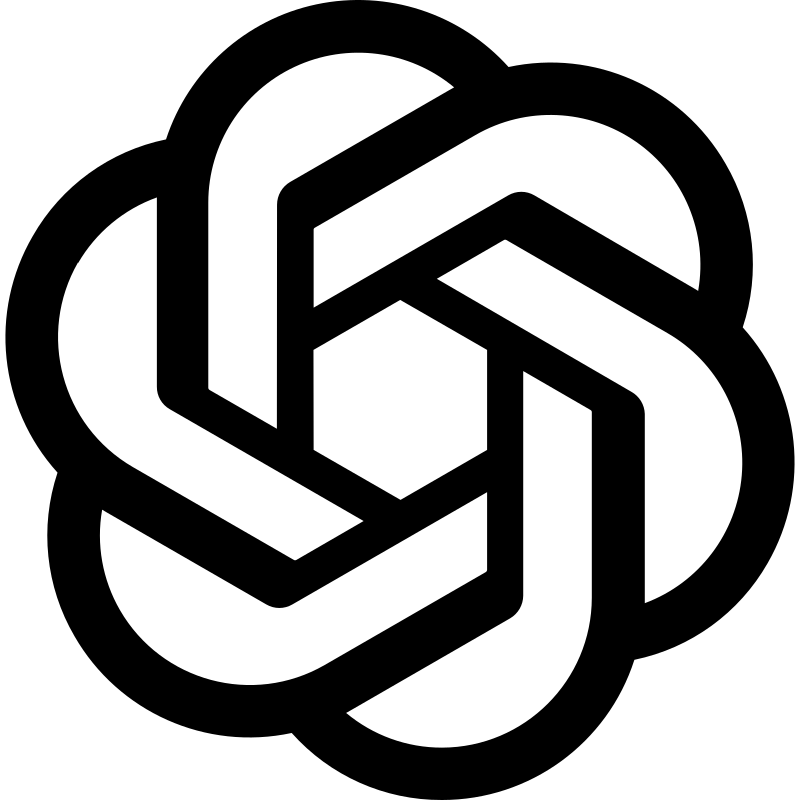}}

\title{SWE-Adept: An LLM-Based Agentic Framework for Deep Codebase Analysis and Structured Issue Resolution}

\author{Kang He \quad Kaushik Roy \\
         Electrical and Computer Engineering, Purdue University \\
         \texttt{\{he603, kaushik\}@purdue.edu}}

\begin{document}
\maketitle
\begin{abstract}
Large language models (LLMs) exhibit strong performance on self-contained programming tasks. However, they still struggle with repository-level software engineering (SWE), which demands (1) deep codebase navigation with effective context management for accurate localization, and (2) systematic approaches for iterative, test-driven code modification to resolve issues.
To address these challenges, we propose SWE-Adept, an LLM-based two-agent framework where a localization agent identifies issue-relevant code locations and a resolution agent implements the corresponding fixes. For issue localization, we introduce agent-directed depth-first search that selectively traverses code dependencies. This minimizes issue-irrelevant content in the agent's context window and improves localization accuracy. 
For issue resolution, we employ adaptive planning and structured problem solving. We equip the agent with specialized tools for progress tracking and Git-based version control. 
These tools interface with a shared working memory that stores code-state checkpoints indexed by execution steps, facilitating precise checkpoint retrieval.
This design enables reliable agent-driven version-control operations for systematic issue resolution, including branching to explore alternative solutions and reverting failed edits.
Experiments on SWE-Bench Lite and SWE-Bench Pro demonstrate that SWE-Adept consistently outperforms prior approaches in both issue localization and resolution, improving the end-to-end resolve rate by up to 4.3\%.\footnote{Our code is available at \url{https://github.com/kang-ml/SWE-Adept}.}
\end{abstract}

\section{Introduction}
\begingroup

Recent advances in large language models (LLMs) have demonstrated remarkable programming capabilities ~\citep{claude-sonnet-4.5, gemini3, gpt-5.2}. However, compared to isolated function- or file-level tasks ~\citep{chen2021evaluating}, resolving real-world software engineering issues is substantially more challenging ~\citep{yang2024sweagent, xia2024agentless}. First, pinpointing the relevant code locations is difficult as code repositories are large and exhibit dense cross-file dependencies. For example, each codebase in SWE-Bench ~\citep{jimenez2024swebench} contains over 3,000 files on average, far exceeding LLM's context limit. More importantly, many issues are not self-contained: identifying the root cause often requires traversing code dependencies while avoiding context-window overflow ~\citep{ouyang2025repograph, yu2025orcaloca}. Second, implementing a correct fix typically requires iterative code modifications and test-driven validations, rather than a one-shot edit ~\citep{zhang2024autocoderover, wang2025ai, wang2025openhands, yang2025survey}.

\begin{figure*}[ht]
\begin{center}
\centerline{\includegraphics[width=\textwidth]{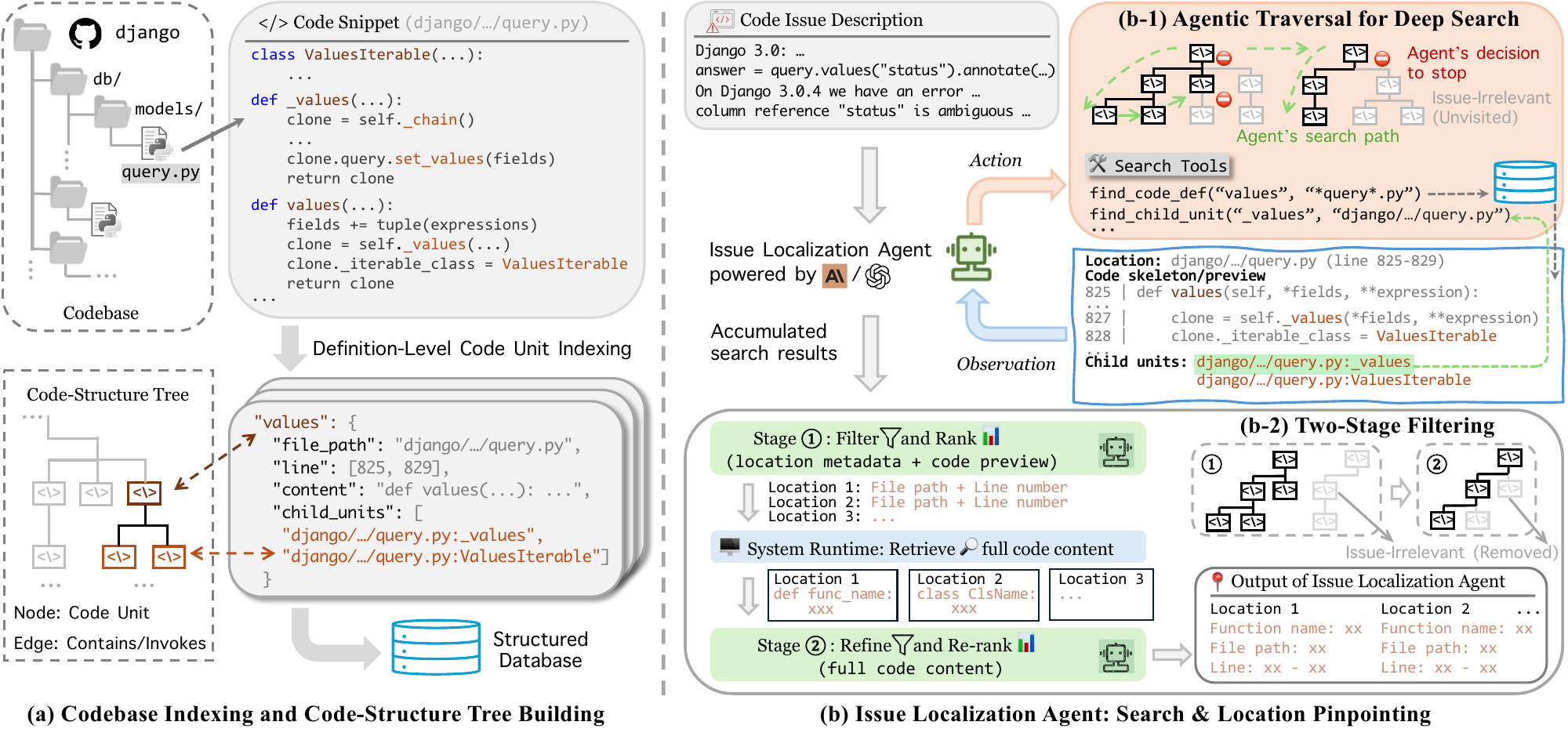}}
\caption{
Overview of \textbf{Issue Localization} framework. (a) Codebase is indexed and represented as code-structure tree in the structured database. Based on this representation, (b) Issue Localization Agent performs search and pinpointing: (b-1) agent-directed depth-first traversal for selective, dependency-aware exploration, with search tools (Table~\ref{localization_tool}) returning lightweight structural information; (b-2) post-search two-stage filtering (code-preview and location heuristics followed by content-based analysis) for candidate re-ranking and final issue-relevant locations.
}
\label{issue_location_agent}
\end{center}
\end{figure*}

\begin{figure*}[ht]
\begin{center}
\centerline{\includegraphics[width=\textwidth]{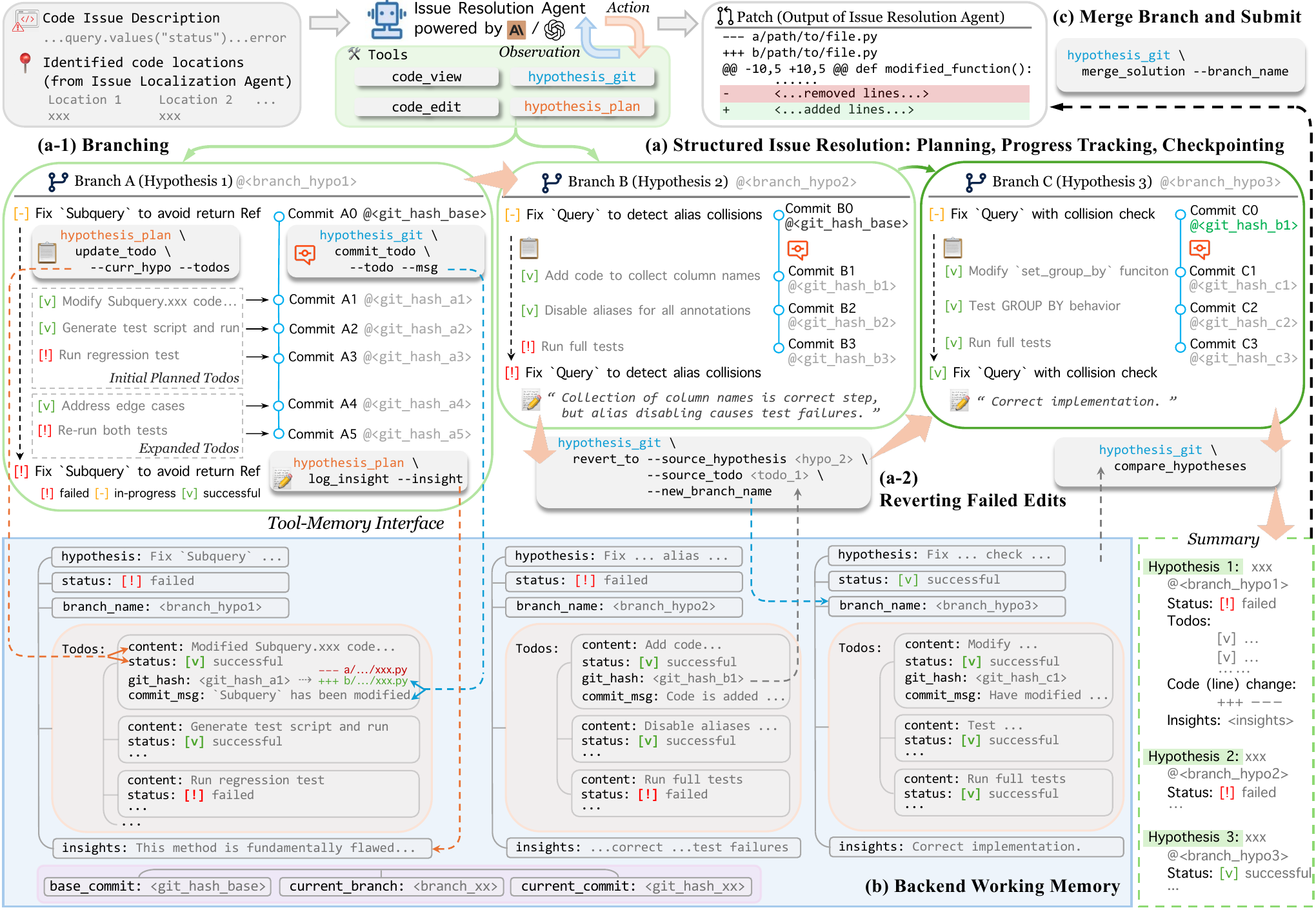}}
\caption{
Overview of \textbf{Issue Resolution} framework. (a) Issue Resolution Agent takes identified code locations as input and performs structured issue resolution. The agent is equipped with two CLI (command-line interface)-based tool families (\S~\ref{sec:issue_resolution}), \texttt{hypothesis\_plan} (Table~\ref{resolution_tool_1}) and \texttt{hypothesis\_git} (Table~\ref{resolution_tool_2}), for planning, progress tracking, and version control. (b) Backend working memory stores structured metadata for hypotheses, to-dos and code-state checkpoints. Both tool families interface with this memory to manage checkpoints for version-control operations, including (a-1) branching to explore alternative solutions (hypotheses) and (a-2) reverting failed edits. (c) The agent merges the selected hypothesis branch after comparing all implemented hypotheses and submits the final patch.
}
\label{issue_resolution_agent}
\end{center}
\end{figure*}

To tackle these challenges, prior research introduces procedure-based pipelines that decompose repository-level debugging into stages consisting of localization, repair, and validation to automate issue resolution ~\citep{xia2024agentless}. Recent work adopts agent-based paradigm, equipping LLMs with tool access and enabling them to iteratively execute actions, observe feedback such as test results, and plan subsequent steps ~\citep{yang2024sweagent, phan2024hyperagent, antoniades2025swesearch, wang2025openhands}. In parallel, to improve issue localization accuracy, several studies build structured codebase representations to support dependency-aware navigation and issue-relevant code retrieval ~\citep{liu-etal-2025-codexgraph, ouyang2025repograph, chen-etal-2025-locagent}. Although these research demonstrate significant advancement, they still exhibit several limitations:

\noindent(1) Less-effective context management during codebase search injects excessive issue-irrelevant information into agent's context, which in turn degrades localization accuracy ~\citep{hsieh2024ruler, liu-etal-2024-lost}. Some approaches employ coarse-grained codebase indexing, causing a single query to yield many candidate matches with insufficient context for effective prioritization ~\citep{yang2024sweagent}. Consequently, the agent pulls in entire files or large spans to disambiguate candidates, quickly consuming the context window with excessive content. Additionally, prior methods use algorithm-controlled traversal with fixed-hop expansion ~\citep{ouyang2025repograph}. This often enforces breadth-first, indiscriminate expansion, introducing redundancy and irrelevant search paths ~\citep{yu2025orcaloca}.

\noindent(2) Existing code-debugging approaches generally lack systematic strategies for issue resolution. Methods such as SWE-agent ~\citep{yang2024sweagent} often operate in a free-form “think-and-edit” loop without explicit planning and progress tracking. As iterations accumulate, interleaved code edits and newly generated test scripts can make the working state difficult to interpret and validate. More importantly, prior approaches rarely include checkpointing mechanism that continuously records intermediate code states aligned with execution milestones. Without such state logging, agents may struggle to reliably revert to a previous intermediate state after failed edits, or reset to the baseline state to attempt alternative repairs.

To address these limitations, we propose SWE-Adept, an LLM-based agentic framework for end-to-end software issue resolution. SWE-Adept consists of two specialized agents: an \textit{issue localization agent} that searches codebase and pinpoints issue-relevant code locations, and an \textit{issue resolution agent} that implements and validates the corresponding fixes.

To enable efficient and precise codebase navigation for issue localization, we first conduct codebase indexing and build code-structure tree from code dependencies, as illustrated in Figure~\ref{issue_location_agent}(a). We then introduce agentic traversal (Figure~\ref{issue_location_agent}(b-1)): the issue localization agent performs \textit{depth-first}, dependency-aware exploration over the tree. During traversal, search tools return only lightweight structural information (code skeleton/preview, invocation context, and location metadata) to minimize context consumption. Once search is complete, the agent follows a two-stage filtering scheme that defers full code content loading to the final re-ranking stage for precise localization, as depicted in Figure~\ref{issue_location_agent}(b-2).

After localization, the issue resolution agent receives the identified code locations and performs structured problem solving (Figure~\ref{issue_resolution_agent}(a)). 
The agent formulates one or more repair hypotheses. For each hypothesis, it creates a fine-grained to-do list and adaptively updates it based on execution feedback.
Furthermore, we design a checkpointing mechanism that captures the intermediate code state after each completed to-do step and stores each checkpoint in a backend working memory, indexed by the corresponding step to enable precise retrieval (Figure~\ref{issue_resolution_agent}(b)).
To achieve this, we equip the agent with specialized tools for progress tracking and Git-based version control. 
These tools interface with the shared working memory to manage code-state checkpoints.
With this checkpointing design, the agent can reliably perform version-control operations, including branching to evaluate alternative solutions (i.e., hypotheses) and reverting failed edits, for systematic long-horizon issue resolution.

Our experimental evaluation on SWE-Bench Lite ~\citep{jimenez2024swebench} and SWE-Bench Pro ~\citep{deng2025swe} demonstrates that SWE-Adept achieves superior performance in both issue localization and issue resolution, improving function-level localization accuracy by up to 6.2\% and end-to-end resolve rate by up to 4.3\%.

The main contributions of our work are:

\begin{itemize}[noitemsep,nolistsep]
\item We propose SWE-Adept, an LLM-based agentic framework that integrates precise issue localization with structured issue resolution to autonomously fix repository-level software engineering issues.

\vspace{2pt}

\item We introduce agentic traversal that enables effective context management through depth-first, dependency-guided codebase navigation, coupled with two-stage filtering to precisely pinpoint issue-relevant code locations.

\vspace{2pt}
\item We design a tool-memory interface that enables reliable agent-driven version control for systematic, long-horizon issue resolution.
\end{itemize}

\endgroup

\section{Related Work}
\begingroup
\subsection{LLM for Software Engineering}
Resolving issues in real-world software systems remains challenging for LLMs. 
To address this, recent work has proposed LLM-based agentic frameworks that empower one or more LLMs with a scaffolding ~\citep{scaffolding} composed of software architecture, tool interface, and prompting strategy, making LLMs more capable on problems that demand deeper code understanding and multi-step problem solving ~\citep{yang2024sweagent, wang2025openhands, jiang2025agentic}. 

In practice, resolving a software issue typically involves two essential subtasks:

\noindent \textbf{Issue localization.}\quad Given an issue description, this subtask aims to identify the code locations (e.g., function, class) that are most likely the root cause and thereby the primary target for editing. Agent-based methods perform localization via multi-step, tool-assisted codebase navigation. SWE-agent ~\citep{yang2024sweagent} introduces an agent-computer interface that enhances agent's ability to search codebase. Graph-based approaches such as LocAgent ~\citep{chen-etal-2025-locagent} and RepoGraph ~\citep{ouyang2025repograph} build dependency graphs over code entities to guide navigation. However, achieving thorough search while maintaining effective context management is challenging yet critical, since excessive irrelevant retrieval can degrade localization accuracy ~\citep{yu2025orcaloca}.

\noindent \textbf{Issue resolution.}\quad Given the locations of target code and relevant context, this subtask generates and applies code changes (i.e., patches) that resolve the reported issue. Agentless ~\citep{xia2024agentless} generates multiple candidate patches and then uses test-based validation and ranking to select the final patch. SWE-agent ~\citep{yang2024sweagent} iteratively edits code and runs tests until it produces a patch that passes all required tests. AutoCodeRover ~\citep{zhang2024autocoderover} employs an LLM-based patching agent and iteratively retries to obtain an applicable patch. OpenHands ~\citep{wang2025openhands} places LLM in a sandbox (e.g., a shell and workspace), enabling command-driven code editing and validation. However, these approaches largely adopt a free-form think-and-act paradigm ~\citep{yao2023react} that can yield a disorganized edit trajectory, where successive modifications accumulate and obscure the causal link between edits and observed outcomes. They provide limited structure for systematic problem solving beyond local trial-and-error ~\citep{yao2023tree, he-roy-2025-logictree}. Some systems, such as SWE-Search~\citep{antoniades2025swesearch} and Claude Code~\citep{claude_code_overview}, support plan-guided execution and code-state checkpointing. However, in SWE-Search, checkpoints are managed by the system runtime and not accessible to the agent through its context or tool interface; in Claude Code, checkpoints appear to be surfaced primarily for user control.
In contrast, our design enables the agent to leverage checkpoints for autonomous version control, supporting systematic, long-horizon issue resolution.

\subsection{Memory for LLM Agents}
Memory increasingly serves as a fundamental component in modern LLM-based agentic framework to support multi-step decision making ~\citep{zhang2025survey, hu2025memory}. HiAgent ~\citep{hu-etal-2025-hiagent} introduces a hierarchical working-memory design that manages intermediate task state for maintaining coherence across extended interaction. MemoryOS ~\citep{kang-etal-2025-memory} models agent memory as an operating-system-like stack with dedicated modules for storing, update, retrieval, and synthesizing information. A-MEM ~\citep{xu2025amem} dynamically integrates prior experiences into a structured graph. In agentic approaches for autonomous software engineering, existing memory designs emphasize reflection and experience reuse to improve agent capability ~\citep{chen2025swe, hayashi2025self}. We take an orthogonal perspective and employ working memory to store code-state checkpoints, facilitating reliable agent-driven version-control operations.

\endgroup

\section{SWE-Adept Framework}
\begingroup
We introduce SWE-Adept, an LLM-based agentic framework for resolving repository-level software engineering issues. Figure~\ref{issue_location_agent} - \ref{issue_resolution_agent} illustrate the end-to-end workflow: codebase indexing and code-structure tree building, issue localization, and issue resolution. SWE-Adept comprises two specialized agents: (i) a localization agent that navigates the repository to identify issue-relevant code locations, and (ii) a resolution agent that conducts and validates the corresponding fixes. The two agents operate in separate context windows with distinct tool access, preventing the full search-and-edit trace from accumulating within a single agent ~\citep{tran2025multi}. In the following sections, we will describe each component of the workflow in detail.

\subsection{Codebase Representation}
Given a code repository $\mathcal{R}$, we construct a fine-grained, definition-level indexing to balance context completeness and context efficiency for downstream codebase navigation. 
We use \texttt{tree-sitter}\footnote{\url{https://tree-sitter.github.io/}} with language-specific grammars to parse and segment repository $\mathcal{R}$ into a set of code units $\mathcal{U}=\{u_i \mid i=1,\ldots,|\mathcal{U}|\}$. We first extract language-level definitions, such as functions and classes, as the primary self-contained units. Remaining code content is segmented into fixed-length chunks of 200 lines and added to $\mathcal{U}$ to ensure full repository coverage.
Each $u_i$ is represented by metadata fields including its name $n(u_i)$ (for function/class units), source location $loc(u_i)=(p,\ell_s,\ell_e)$ (file path and start/end line numbers), and raw code text $code(u_i)$. This index enables precise referencing of target code units without loading the full file or large surrounding spans. Furthermore, similar to OrcaLoca ~\citep{yu2025orcaloca} and LocAgent ~\citep{chen-etal-2025-locagent}, we leverage dependencies between code units to construct a code-structure tree $\mathcal{T}=(\mathcal{V}, \mathcal{E})$ that facilitates dependency-aware codebase navigation (Figure~\ref{issue_location_agent}(a)). Each node $v \in \mathcal{V}$ corresponds to a code unit $u_i$ extracted during indexing, and each directed edge $e \in \mathcal{E}$ represents \texttt{contains} or \texttt{invokes}, defining a parent-child relation between units. A key difference of our representation is that, rather than constructing a monolithic global code graph, we store a lightweight adjacency list as part of each code unit's metadata. For $u_i$, its adjacency list $adj(u_i)$ contains the identifiers \colorbox{codegray}{\texttt{file\_path:definition\_name}} of its child units. This design is more retrieval-efficient: accessing a unit returns both the unit and its local adjacency, avoiding a separate dependency lookup and reducing traversal overhead.

\subsection{Issue Localization}
\label{sec:localization}
\noindent \textbf{Tool design.}\quad 
Table~\ref{localization_tool} lists the tools provided to Issue Localization Agent. The tool set supports search across multiple granularities: file-level retrieval, class/function-level definition lookup, and line/variable-level content matching. In addition to these search capabilities, we include \texttt{find\_child\_unit} to make codebase navigation explicit in the agent's action space.
Across all search tools, the minimum retrieval unit is an indexed code unit $u_i$. 
To improve context efficiency, the tools return concise structural information instead of full source content. Each result includes location metadata (file path and line span) and provides file skeleton (for file-level retrieval) or class/function preview with child-unit identifiers (Figure~\ref{issue_location_agent}(b)). We also include \texttt{finish\_search} to signal the completion of search.

\noindent \textbf{Localization agent operation.}\quad
We introduce agent-directed depth-first traversal strategy that performs selective exploration over dependency paths.
To start, the localization agent extracts code entities explicitly mentioned in the issue description (e.g., file or function names) as initial entry-point keywords to invoke the search tools. If exact code references are unavailable, the agent performs pattern-based search (e.g., partial-string queries) and the tools return a ranked list of candidate code units (Table~\ref{localization_tool}). Based on the tool outputs (code skeleton/preview with child-unit identifiers), the agent prioritizes one child unit at each step for deep exploration via \texttt{find\_child\_unit}. It recursively applies this action, following a single dependency path that is most likely issue-related. Exploration along the current path stops once the agent has sufficient understanding or determines that the path is issue-unrelated, after which it moves to the next candidate path or entry point (Figure~\ref{issue_location_agent}(b-1)). Once the agent has obtained sufficient context, it invokes \texttt{finish\_search} to end the search phase.

After exploration, the agent conducts a two-stage filtering as shown in Figure~\ref{issue_location_agent}(b-2). The first stage shortlists candidate locations using lightweight heuristics (code skeleton/preview with invocation context, location metadata), removing clearly issue-irrelevant exploration paths. The system runtime then retrieves the full source code for the shortlisted locations and provides it as input for the second stage. 
This deferred full code loading minimizes redundant retrieval.
In the second stage, the agent analyzes the full code implementation to further refine and re-rank the candidate locations. 

\subsection{Issue Resolution}
\label{sec:issue_resolution}
We build Issue Resolution Agent on SWE-agent ~\citep{yang2024sweagent}, leveraging its infrastructure while adapting its workflow to incorporate our proposed tool families and backend working memory.

\noindent \textbf{Tool design.}\quad 
We introduce two CLI (command-line interface)-based tool families for issue resolution: \texttt{hypothesis\_plan} (Table~\ref{resolution_tool_1}) and \texttt{hypothesis\_git} (Table~\ref{resolution_tool_2}). Each family comprises multiple related commands under a common namespace. \texttt{hypothesis\_plan} maintains (i) a set of hypotheses (i.e., alternative solutions) and (ii) hypothesis-associated to-do lists, and tracks their execution status. It also logs insights obtained from execution feedback. \texttt{hypothesis\_git} conducts Git-based version-control operations, including branching and commit-based checkpointing.
Each \texttt{hypothesis\_git} command wraps a sequence of low-level Git operations into a single high-level action with built-in error handling. This abstraction minimizes version-control mistakes, since multi-step Git workflows executed directly by the agent are highly error-prone over long trajectories.

Both tool families interface with a shared working memory that stores the associations among hypotheses, their to-do steps, and the corresponding checkpoint metadata (Git hashes and commit messages), as shown in Figure~\ref{issue_resolution_agent}(b). 
The agent invokes tools using semantic identifiers (e.g., hypothesis and to-do names) as arguments; the tools access working memory to store and retrieve the associated code-state information (e.g., branch names and Git hashes).
This removes the requirement for the agent to track non-semantic Git hashes in-context and enables reliable code-state management, especially under heavy branching and checkpointing.

\noindent \textbf{Resolution agent operation.}\quad 
The resolution agent receives the identified code locations from the localization agent and uses these anchors to initialize its analysis. It first invokes \texttt{hypothesis\_git} to checkpoint the original code state, then generates and runs a reproduction script to confirm the reported issue. It next performs hypothesis-driven repair. For complex issues (e.g., when the fix spans multiple code locations or involves intricate dependencies), it formulates and evaluates multiple competing hypotheses; otherwise, it proceeds with a single hypothesis when the root cause and fix strategy are clear. The agent explores one hypothesis at a time. For each hypothesis, the agent checks out an isolated branch and initializes a to-do plan of fine-grained \textit{edit} and \textit{test} actions. Planning is adaptive. If test feedback reveals uncovered edge cases or missing steps, the agent adds new to-do items as needed, as illustrated in Figure~\ref{issue_resolution_agent} Branch A.

Execution is checkpointed step-by-step. After each to-do, the agent invokes \texttt{hypothesis\_git} to commit the current state as a code-state checkpoint. This invocation automatically stores the checkpoint metadata (Git hash and commit message) in the working memory and links it to the completed step. This semantic-step indexing of checkpoints facilitates reliable version-control operations for systematic problem solving. When a hypothesis proves partially correct (i.e., earlier steps remain useful but later direction is wrong), the agent reverts edits from the failed later steps by returning to the appropriate prior checkpoint using the semantic-step reference. By design, it then checks out a new branch to continue exploration, keeping alternative solution trajectories cleanly separated, as demonstrated in Figure~\ref{issue_resolution_agent}(a-2).

After exploring all hypotheses, the agent invokes \texttt{hypothesis\_git} to compile a comparative report across hypotheses, summarizing their status, to-dos, commits, code diffs, and insights to support final selection. It then merges the selected hypothesis branch into the checkpoint of original code state for submission, as shown in Figure~\ref{issue_resolution_agent}(c).

\endgroup

\section{Experiments}
\begingroup

\subsection{Experimental Setup}
\label{metric}
\begingroup

\begin{table*}
\small
\centering
\setlength{\tabcolsep}{4pt}
\renewcommand{\arraystretch}{1.1}

\newcolumntype{L}{>{\raggedright\arraybackslash}X} 
\newcolumntype{C}{>{\centering\arraybackslash}X}  

\begin{tabularx}{\textwidth}{LLcccccc}
\toprule
\multirow{2.5}{*}{Framework} 
  & \multirow{2.5}{*}{Model} 
  & \multicolumn{3}{c}{SWE-Bench Lite} 
  & \multicolumn{3}{c}{SWE-Bench Pro} \\
\cmidrule(lr){3-5}\cmidrule(lr){6-8}
  & 
  & File Acc@3
  & Func Acc@5 
  & \# Tokens
  & File Acc@3 
  & Func Acc@5 
  & \# Tokens \\
\midrule

\multirow[t]{2}{*}{Embedding}
  & CodeSage-Large
  & 71.2\%
  & 40.1\%
  & N/A
  & 54.0\%
  & 26.5\%
  & N/A\\

\multirow[t]{2}{*}{Embedding}
  & CodeRankEmbed
  & 76.6\%
  & 50.0\%
  & N/A
  & 61.0\%
  & 33.0\%
  & N/A\\

\midrule

\rowcolor[HTML]{f2f2f2}
\multirow[t]{2}{*}{SWE-agent}
  & \openai{} GPT-5.2
  & 84.3\%
  & 60.0\%
  & 78k
  & 71.0\%
  & 37.8\%
  & 125k\\
\rowcolor[HTML]{f2f2f2}
  & \anthropic{} Claude-4.5
  & 90.5\%
  & 78.6\%
  & 298k
  & 78.0\%
  & 55.0\%
  & 340k\\

\multirow[t]{2}{*}{RepoGraph}
  & \openai{} GPT-5.2
  & 86.8\%
  & 60.6\%
  & 150k
  & 71.5\%
  & 40.0\%
  & 272k\\
  & \anthropic{} Claude-4.5
  & 92.0\%
  & 79.2\%
  & 412k
  & 80.0\%
  & 56.3\%
  & 473k\\

\rowcolor[HTML]{f2f2f2}
\multirow[t]{2}{*}{OrcaLoca}
  & \openai{} GPT-5.2
  & 87.2\%
  & 64.2\%
  & 265k
  & \underline{74.5\%}
  & \underline{43.0\%}
  & 368k\\
\rowcolor[HTML]{f2f2f2}
  & \anthropic{} Claude-4.5
  & \underline{94.7\%}
  & \underline{83.4\%}
  & 468k
  & 81.5\%
  & \underline{57.5\%}
  & 518k\\

\multirow[t]{2}{*}{LocAgent}
  & \openai{} GPT-5.2
  & \underline{87.5\%}
  & \underline{64.6\%}
  & 215k
  & 73.8\%
  & 42.3\%
  & 355k\\
  & \anthropic{} Claude-4.5
  & 93.5\%
  & 81.8\%
  & 380k
  & \underline{82.0\%}
  & 57.3\%
  & 450k\\

\midrule

\rowcolor[HTML]{f2f2f2}
\multirow[t]{2}{*}{\textbf{SWE-Adept}}
  & \openai{} GPT-5.2
  & \textbf{92.3\%}
  & \textbf{70.8\%}
  & 197k
  & \textbf{77.5\%}
  & \textbf{46.0\%}
  & 252k\\
\rowcolor[HTML]{f2f2f2}
  & \anthropic{} Claude-4.5
  & \textbf{97.0\%}
  & \textbf{87.8\%}
  & 348k
  & \textbf{85.8\%}
  & \textbf{63.3\%}
  & 427k\\

\bottomrule
\end{tabularx}
\caption{Issue localization performance of different frameworks on SWE-Bench Lite and SWE-Bench Pro with GPT-5.2 and Claude-Sonnet-4.5 (Claude-4.5). Accuracy is reported at file and function levels (\S~\ref{metric}). Best results are in bold and second-best is underlined. \# Tokens denotes the total number of input and output tokens per instance.}
\label{localization_result}
\end{table*}

\endgroup

\noindent \textbf{Datasets.}\quad 
We evaluate our framework on two repository-level software engineering benchmarks: SWE-Bench Lite ~\citep{jimenez2024swebench} and SWE-Bench Pro ~\citep{deng2025swe}. Each instance is curated from a real-world GitHub issue and its associated codebase. The task is to submit a patch that edits the relevant code to resolve the issue. Additional dataset details are provided in Appendix~\ref{dataset}.

\noindent \textbf{Metrics.}\quad We evaluate performance on both issue localization and issue resolution.
\begin{itemize}[leftmargin=10pt,noitemsep,nolistsep]
    \vspace{1pt}
    \item \textit{Issue localization.} We follow ~\citet{chen-etal-2025-locagent} and use Acc@$k$. For each instance, the localization agent outputs a ranked list of locations. We take the top-$k$ predictions and mark the instance as correct only if all locations modified in the ground-truth patch are contained in the top-$k$ set. We report Acc@3 for file-level localization and Acc@5 for function-level localization.
    \vspace{1pt}
    \item \textit{Issue resolution.} We report the resolve rate ~\citep{jimenez2024swebench}, defined as the percentage of instances successfully resolved over the dataset. An instance counts as resolved if the submitted patch passes all corresponding tests.
\end{itemize}

\noindent \textbf{Baselines.}\quad We compare our framework with representative baselines for both issue localization and end-to-end issue resolution. 

\begin{itemize}[leftmargin=10pt,noitemsep,nolistsep]
\vspace{1pt}
\item \textit{Issue localization}: (1) Embedding-based retrieval: CodeSage-Large, a 1.3B encoder model ~\citep{zhang2024code}; CodeRankEmbed, a 137M encoder model ~\citep{suresh2025cornstack}. (2) LLM-based agentic localization: SWE-agent uses an agent-computer interface for code search ~\citep{yang2024sweagent}; RepoGraph ~\cite{ouyang2025repograph} and LocAgent ~\citep{chen-etal-2025-locagent} build graph representations of the codebase to support navigation; OrcaLoca ~\citep{yu2025orcaloca} designs specialized sub-agents to improve localization.
\vspace{1pt}
\item \textit{Issue resolution}: SWE-agent; RepoGraph (integrated with SWE-agent for patch generation); OpenHands ~\citep{wang2025openhands}; SWE-Search ~\citep{antoniades2025swesearch}; OrcaLoca (integrated with Agentless for patch generation).
\end{itemize}

We separately employ GPT-5.2 and Claude-Sonnet-4.5 model in our framework, and reproduce other agentic approaches using the same models for comparison. We additionally compare with Claude Code~\citep{claude_code_overview} and mini-SWE-Agent\footnote{\url{https://mini-swe-agent.com/latest/}}. Beyond proprietary models, we also evaluate SWE-Adept with two open-source models, GLM-4.7-Flash (30B)~\citep{GLM-4.7} and Qwen3.5-27B~\citep{Qwen3.5}, following the same experimental settings. The corresponding results and implementation details are provided in Appendix~\ref{model}.

\begingroup

\begin{table*}
\small
\centering
\setlength{\tabcolsep}{4pt}
\renewcommand{\arraystretch}{1.1}

\newcolumntype{Y}{>{\centering\arraybackslash}X}

\begin{tabularx}{\textwidth}{ l >{\hspace{2.8em}}l<{\hspace{2em}} YYYY }
\toprule
\multirow{2.5}{*}{Framework} 
  & \multirow{2.5}{*}{Model} 
  & \multicolumn{2}{c}{SWE-Bench Lite} 
  & \multicolumn{2}{c}{SWE-Bench Pro} \\
\cmidrule(lr){3-4}\cmidrule(lr){5-6}
  & 
  & Resolve Rate
  & \# Tokens
  & Resolve Rate
  & \# Tokens \\

\midrule

\multirow[t]{2}{*}{SWE-agent}
  & \openai{} GPT-5.2
  & 52.5\%
  & 260k
  & 30.3\%
  & 338k\\
  & \anthropic{} Claude-4.5
  & 66.0\%
  & 2640k
  & 39.5\%
  & 3320k\\

\rowcolor[HTML]{f2f2f2}
\multirow[t]{2}{*}{RepoGraph}
  & \openai{} GPT-5.2
  & 50.3\%
  & 272k
  & 30.0\%
  & 431k\\
\rowcolor[HTML]{f2f2f2}
  & \anthropic{} Claude-4.5
  & 66.2\%
  & 2968k
  & 38.0\%
  & 3815k\\

\multirow[t]{2}{*}{OpenHands}
  & \openai{} GPT-5.2
  & 51.5\%
  & 275k
  & 29.5\%
  & 365k\\
  & \anthropic{} Claude-4.5
  & 65.2\%
  & 2680k
  & 40.0\%
  & 3506k\\

\rowcolor[HTML]{f2f2f2}
\multirow[t]{2}{*}{SWE-Search}
  & \openai{} GPT-5.2
  & 54.5\%
  & 820k
  & 32.5\%
  & 1019k\\
\rowcolor[HTML]{f2f2f2}
  & \anthropic{} Claude-4.5
  & 66.8\%
  & 3368k
  & 40.8\%
  & 4380k\\

\multirow[t]{2}{*}{OrcaLoca}
  & \openai{} GPT-5.2
  & \underline{56.0\%}
  & 495k
  & \underline{33.5\%}
  & 694k\\
  & \anthropic{} Claude-4.5
  & \underline{68.3\%}
  & 1875k
  & \underline{41.3\%}
  & 2850k\\

\midrule

\rowcolor[HTML]{f2f2f2}
\multirow[t]{2}{*}{\textbf{SWE-Adept}}
  & \openai{} GPT-5.2
  & \textbf{58.8\%}
  & 703k
  & \textbf{37.8\%}
  & 864k\\
\rowcolor[HTML]{f2f2f2}
  & \anthropic{} Claude-4.5
  & \textbf{71.3\%}
  & 3119k
  & \textbf{45.0\%}
  & 4085k\\

\bottomrule
\end{tabularx}
\caption{End-to-end issue resolve rate of different frameworks on SWE-Bench Lite and SWE-Bench Pro with GPT-5.2 and Claude-Sonnet-4.5 (Claude-4.5). Best results are in bold and second-best is underlined. \# Tokens denotes the total number of input and output tokens per instance, including issue localization and issue resolution.}
\label{resolution_result}
\end{table*}

\endgroup

\subsection{Main Results}
As shown in Table~\ref{localization_result} and Table~\ref{resolution_result}, our framework consistently outperforms baseline approaches on both issue localization and issue resolution. For localization accuracy at function level (Func Acc@5), on SWE-Bench Lite, it improves over the strongest baseline by 6.2\% with GPT-5.2 and 4.4\% with Claude-Sonnet-4.5; on SWE-Bench Pro, the corresponding gains are 3.0\% and 5.8\%, respectively. In addition, our framework mostly consumes fewer tokens than graph-based approaches (RepoGraph, OrcaLoca, LocAgent) due to effective context management (\S~\ref{sec:localization}). For end-to-end issue resolution, on SWE-Bench Lite, our framework achieves 2.8\% and 3.0\% higher resolve rate with GPT-5.2 and Claude-Sonnet-4.5, respectively; on SWE-Bench Pro, the corresponding improvements are 4.3\% and 3.7\%. For SWE-Bench Pro, we further report localization and resolution results by programming language in Table~\ref{language_wise_result}. SWE-Adept achieves stronger performance across all the evaluated programming languages, demonstrating its broad applicability.
\endgroup

\section{Further Analysis}
\begingroup
\subsection{Agent Action Patterns and Performance}

Figure~\ref{localization_behavior}(a) presents the distribution of search actions invoked by Issue Localization Agent. The most frequent action \texttt{find\_child\_unit} indicates that the agent primarily performs dependency-aware multi-hop navigation over the code-structure tree. 
For each instance, we measure the maximum search depth across all explored paths in the agent's trajectory, and report the instance distribution and localization accuracy by maximum search depth in Figure~\ref{localization_behavior}(b).
Localization accuracy increases from zero search depth (i.e., no \texttt{find\_child\_unit} call) to moderate search depth, highlighting the importance of deep codebase exploration for identifying root cause. 
Localization accuracy decreases at higher search depth, indicating greater problem difficulty.
Despite this, for instances with search depth greater than zero, SWE-Adept consistently outperforms SWE-agent and OrcaLoca, and its advantage over OrcaLoca becomes more pronounced as search depth increases.
This gain comes from more effective context management which minimizes issue-irrelevant context during search.

Figure~\ref{resolution_behavior}(a) highlights three problem-solving behaviors of Issue Resolution Agent. The high frequency of \textit{multi-hypothesis branching} indicates that the agent often explores multiple candidate solutions, which is beneficial for complex issues. \textit{Dynamic to-do expansion} shows that the agent adaptively updates its plan during execution based on implementation feedback. The observed \textit{checkpoint-based reversion} demonstrates the agent's capability to revert incorrect code changes during iterative problem-solving. All these behaviors are achieved by agent-driven version-control operations through \texttt{hypothesis\_git} invocations, with the tool accessing working memory to store and retrieve code-state checkpoints. In Figure~\ref{resolution_behavior}(b), we plot the instance distribution and resolve rate by number of explored hypotheses.
Although resolve rate declines as hypothesis count increases, indicating higher task complexity, our framework shows better robustness and consistently outperforms other approaches.
We provide additional error analysis in Appendix~\ref{error_analysis}.

\subsection{Ablation Study}

We evaluate the contribution of each agent in our framework. The results in Table~\ref{resolve_rate_ablation} show that SWE-Adept's overall advantage arises from the combination of accurate localization and systematic issue resolution.
For issue localization, we compare our proposed context-management design against a baseline that directly returns complete source code during search (Table~\ref{search_ablation}). Using compact code previews and specialized filtering reduces token consumption while improving localization accuracy. This demonstrates that minimizing issue-irrelevant context enables more precise localization. 
For issue resolution, we evaluate whether the agent can reliably manage code states using raw Git commands. As shown in Table~\ref{raw_git_ablation}, direct raw-Git usage does not reproduce the gains of our method. It improves over vanilla SWE-agent on SWE-Bench Lite, but degrades on the more challenging SWE-Bench Pro.
The main issue is long-horizon reliability: continuous Git operations executed directly by the agent are error-prone over long trajectories, and the growing number of checkpoints makes code-state tracking harder as context accumulates. In our framework, \texttt{hypothesis\_git} wraps low-level Git commands into higher-level actions with built-in error handling. Furthermore, it interfaces with working memory to store and retrieve code-state checkpoints indexed by semantic execution steps. This enables more reliable code-state management for systematic, long-horizon issue resolution.

\endgroup

\section{Conclusion}
\begingroup
We present SWE-Adept, an LLM-based agentic framework for resolving software engineering issues. SWE-Adept comprises two specialized agents, dedicated to issue localization and issue resolution, respectively. For issue localization, we introduce agent-directed depth-first traversal followed by two-stage filtering. The proposed approach enables deep codebase analysis with effective context management, leading to more precise issue localization.
For issue resolution, we employ code-state checkpointing and design a tool-memory interface for code-state management in long-horizon settings. This design enables reliable agent-driven version-control operations for systematic problem solving. Experimental results show that the joint enhancement in issue localization and issue resolution yields superior overall performance for SWE-Adept, highlighting its strength in autonomous software engineering.

\endgroup

\section*{Limitations}
Our evaluation primarily focuses on proprietary LLMs (GPT and Claude models), which demonstrate strong coding performance and robust agentic behavior, including instruction following and tool use in long-horizon tasks. We additionally validate SWE-Adept on two open-source models, where our framework yields consistent gains, though absolute performance still trails the proprietary models. Closing this gap, for example through agentic reinforcement learning, is a promising direction.

Additionally, SWE-Adept does not yet support self-evolution across tasks. Each issue is solved independently, without automatically accumulating debugging knowledge from prior runs. Future work could incorporate experience-driven skill learning to enable continuous improvement over software engineering tasks.

\section*{Ethical Considerations}
Our research complies with the Code of Ethics. We properly cite all models, methods, and datasets used in this work. The benchmark datasets in our experiments are publicly available, and our study does not use private or sensitive data. Our use of datasets and LLMs is consistent with their licenses, terms, and intended usage. Our framework presents some potential risks: as with any autonomous code generation system, SWE-Adept may produce incorrect patches, which could introduce errors if deployed without strict review and testing; and the use of proprietary LLMs may raise privacy concerns. Nevertheless, with proper supervision, our framework can improve the reliability and efficiency of software engineering.

\section*{Acknowledgments}
This work was supported in part by the Center for Co-Design of Cognitive Systems (CoCoSys), a Semiconductor Research Corporation (SRC) and DARPA-sponsored JUMP 2.0 center.

\bibliography{custom}


\clearpage
\appendix

\begingroup
\section{Experimental Details}
\subsection{Datasets}
\label{dataset}
Details of the evaluation datasets are as follows:

\noindent\textbf{SWE-Bench Lite}, a subset of \textbf{SWE-Bench} ~\citep{jimenez2024swebench}, contains 300 instances from 11 GitHub repositories in Python. For function-level localization, we follow \citet{chen-etal-2025-locagent} and exclude instances whose ground-truth patches do not modify any existing functions, retaining 274 instances. For file-level localization and resolve rate, we report results on all 300 instances.

\noindent\textbf{SWE-Bench Pro} ~\citep{deng2025swe} features higher problem complexity, often requiring edits that span multiple files or functions. In our experiments, we evaluate on a 200-instance subset of SWE-Bench Pro test set to control computational cost. To ensure language coverage, we randomly sample 60 instances each from Python, JavaScript, and Go, and include all 20 TypeScript instances available in the public test set. We report issue localization and resolution performance by programming language in Table~\ref{language_wise_result}.

\subsection{Implementation Details}
\label{model}
Here are the versions of GPT-5.2~\citep{gpt-5.2} and Claude-Sonnet-4.5 ~\citep{claude-sonnet-4.5} model:

\noindent \texttt{gpt-5.2-2025-12-11} (medium)

\noindent \texttt{claude-sonnet-4-5-20250929}

\noindent Both models are accessed via API. 

Codebase indexing, code-structure tree construction, and execution of Issue Localization Agent do not require Docker. Building the index and code-structure tree for a repository takes less than one minute, making re-indexing low-overhead when the codebase changes. The ranked location predictions from Issue Localization Agent are stored in a local file and used as input to Issue Resolution Agent. To evaluate the correctness of the patch generated by Issue Resolution Agent, we launch a Docker container for each instance (following the SWE-agent evaluation setup\footnote{\url{https://www.swebench.com/SWE-bench/}}), apply the patch, and execute the tests. Working memory is represented as a persistent, JSON-serialized state structure stored in a shared registry.

We implement Issue Localization Agent using LiteLLM\footnote{\url{https://www.litellm.ai/}} library, and we build Issue Resolution Agent on SWE-agent to leverage its infrastructure. We apply prompt caching\footnote{\url{https://platform.claude.com/docs/en/build-with-claude/prompt-caching}} to both agents to reduce API cost. We set the temperature to 0.1 for both models and report results averaged over two runs. For issue localization, the maximum number of iterations is set to 20 for each instance. For issue resolution, the per-instance cost limit is set to \$5. Under these settings, our framework costs \$1.79 per instance with Claude-Sonnet-4.5 and \$0.42 per instance with GPT-5.2. 
Additionally, we evaluate Claude Code (\texttt{claude-agent-sdk~0.2.87}) and mini-SWE-Agent (bash-only tool access) with Claude-Sonnet-4.5 as comparison baselines under the same settings in Table~\ref{compare_with_cc}.
Experiments with GLM-4.7-Flash (30B) and Qwen3.5-27B on SWE-Bench Lite and SWE-Bench Pro are reported in Table~\ref{open_source_models}.

Prompt for Issue Localization Agent is shown in Figure~\ref{localization_agent_prompt}, and its tools are listed in Table~\ref{localization_tool}. Prompt for Issue Resolution Agent is shown in Figure~\ref{resolution_agent_prompt}, and its tools are listed in Table~\ref{resolution_tool_1} and Table~\ref{resolution_tool_2}.

\section{Error Analysis}
\label{error_analysis}
To analyze failure modes in unresolved instances, we manually review the instances uniquely failed by SWE-agent and SWE-Adept, and group them into three categories (Figure~\ref{venn_diagram}): failure to recover from incorrect edits (\textit{failed recovery}), incorrect solution direction (\textit{incorrect hypothesis}), and wrong function localization (\textit{localization error}). Compared with SWE-agent, SWE-Adept reduces failures across all three categories. This demonstrates that SWE-Adept's performance gains come from the joint contribution of accurate issue localization and systematic issue resolution.

\begin{figure}[ht]
\centering
\includegraphics[width=1\linewidth]{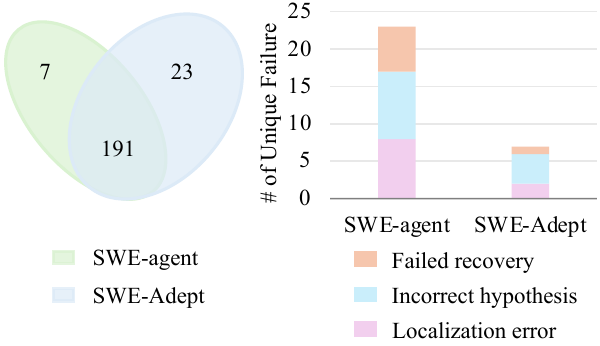}
\caption{Venn diagram (left) of resolved-instance overlap between SWE-Adept and SWE-agent; and error breakdowns (right) for instances uniquely failed by each method. Reported results are on SWE-Bench Lite with Claude-Sonnet-4.5.}
\label{venn_diagram}
\end{figure}

\begingroup

\begin{table*}[t]
\small
\centering
\setlength{\tabcolsep}{2.5pt}
\renewcommand{\arraystretch}{1.1}

\newcolumntype{Y}{>{\centering\arraybackslash}X}

\begin{tabularx}{\textwidth}{l >{\hspace{1.0em}}l<{\hspace{0.8em}} *{8}{Y}}
\toprule
\multirow{2.5}{*}{Framework} 
  & \multirow{2.5}{*}{Model} 
  & \multicolumn{2}{c}{Python} 
  & \multicolumn{2}{c}{JavaScript}
  & \multicolumn{2}{c}{Go}
  & \multicolumn{2}{c}{TypeScript} \\
\cmidrule(lr){3-4}\cmidrule(lr){5-6}\cmidrule(lr){7-8}\cmidrule(lr){9-10}
  &
  & \textit{Localize}
  & \textit{Resolve}
  & \textit{Localize}
  & \textit{Resolve}
  & \textit{Localize}
  & \textit{Resolve}
  & \textit{Localize}
  & \textit{Resolve} \\

\midrule

\rowcolor[HTML]{f2f2f2}
\multirow[t]{2}{*}{SWE-agent}
  & \openai{} GPT-5.2
  & 43.3\% & 38.3\%
  & 40.0\% & 29.2\%
  & 30.8\% & 24.2\%
  & 35.0\% & 27.5\%\\
\rowcolor[HTML]{f2f2f2}
  & \anthropic{} Claude-4.5
  & 61.7\% & 51.7\%
  & 54.2\% & 37.5\%
  & 50.8\% & 30.8\%
  & 50.0\% & 35.0\%\\

\multirow[t]{2}{*}{OrcaLoca}
  & \openai{} GPT-5.2
  & 50.0\% & 42.5\%
  & 45.8\% & 34.2\%
  & 34.2\% & 25.0\%
  & 40.0\% & 30.0\%\\
  & \anthropic{} Claude-4.5
  & 64.2\% & 53.3\%
  & 56.7\% & 40.0\%
  & 53.3\% & 31.7\%
  & 52.5\% & 37.5\%\\

\midrule

\rowcolor[HTML]{f2f2f2}
\multirow[t]{2}{*}{\textbf{SWE-Adept}}
  & \openai{} GPT-5.2
  & \textbf{55.0\%} & \textbf{48.3\%}
  & \textbf{47.5\%} & \textbf{37.5\%}
  & \textbf{36.7\%} & \textbf{26.7\%}
  & \textbf{42.5\%} & \textbf{40.0\%}\\
\rowcolor[HTML]{f2f2f2}
  & \anthropic{} Claude-4.5
  & \textbf{70.0\%} & \textbf{58.3\%}
  & \textbf{63.3\%} & \textbf{44.2\%}
  & \textbf{58.3\%} & \textbf{32.5\%}
  & \textbf{57.5\%} & \textbf{45.0\%}\\

\bottomrule
\end{tabularx}

\caption{Issue localization and resolution performance by programming language on SWE-Bench Pro subset with GPT-5.2 and Claude-Sonnet-4.5 (Claude-4.5). The subset contains 60 Python, 60 JavaScript, 60 Go, and 20 TypeScript instances. \textit{Localize} denotes function-level localization accuracy (Func Acc@5), and \textit{Resolve} denotes end-to-end issue resolve rate.}
\label{language_wise_result}
\end{table*}

\endgroup

\begin{figure*}[t]
\centering
\includegraphics[width=1\textwidth]{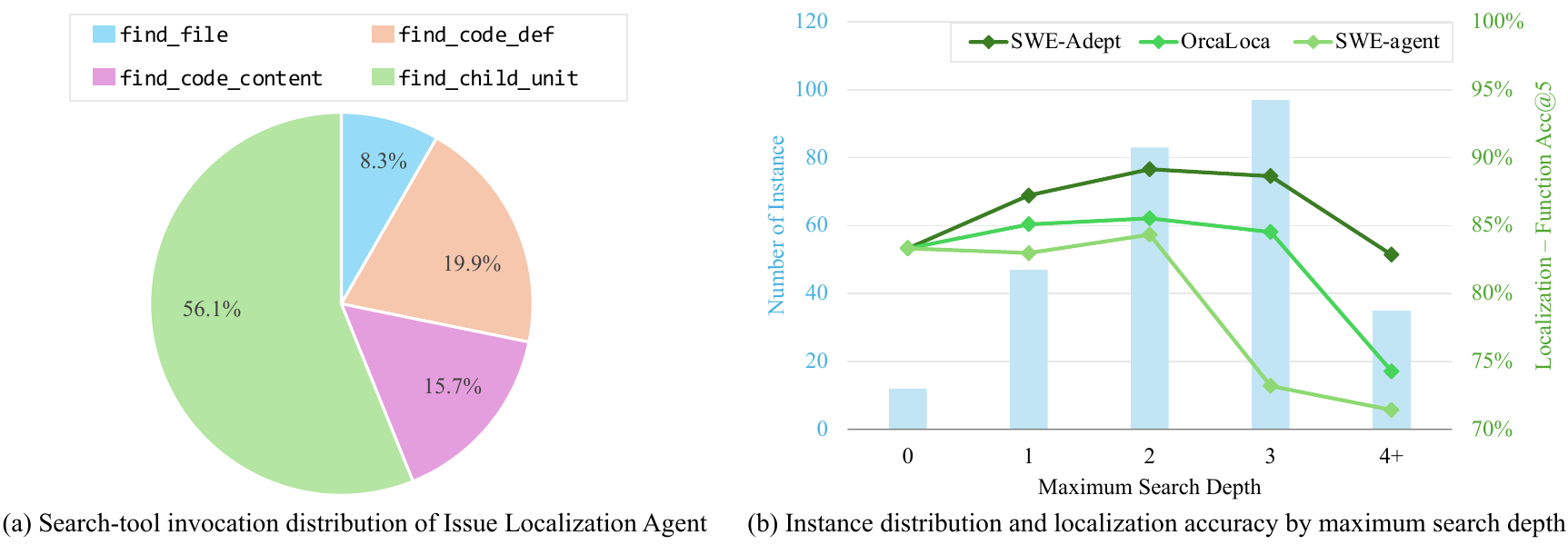}
\caption{Search behavior of \textbf{Issue Localization Agent} and localization accuracy by maximum search depth. (a) Search-tool invocation distribution of Issue Localization Agent. (b) Instance distribution (bars) and function-level localization accuracy (lines) by maximum search depth on SWE-Bench Lite with Claude-Sonnet-4.5.}
\label{localization_behavior}
\end{figure*}

\begin{figure*}[t]
\centering
\includegraphics[width=1\textwidth]{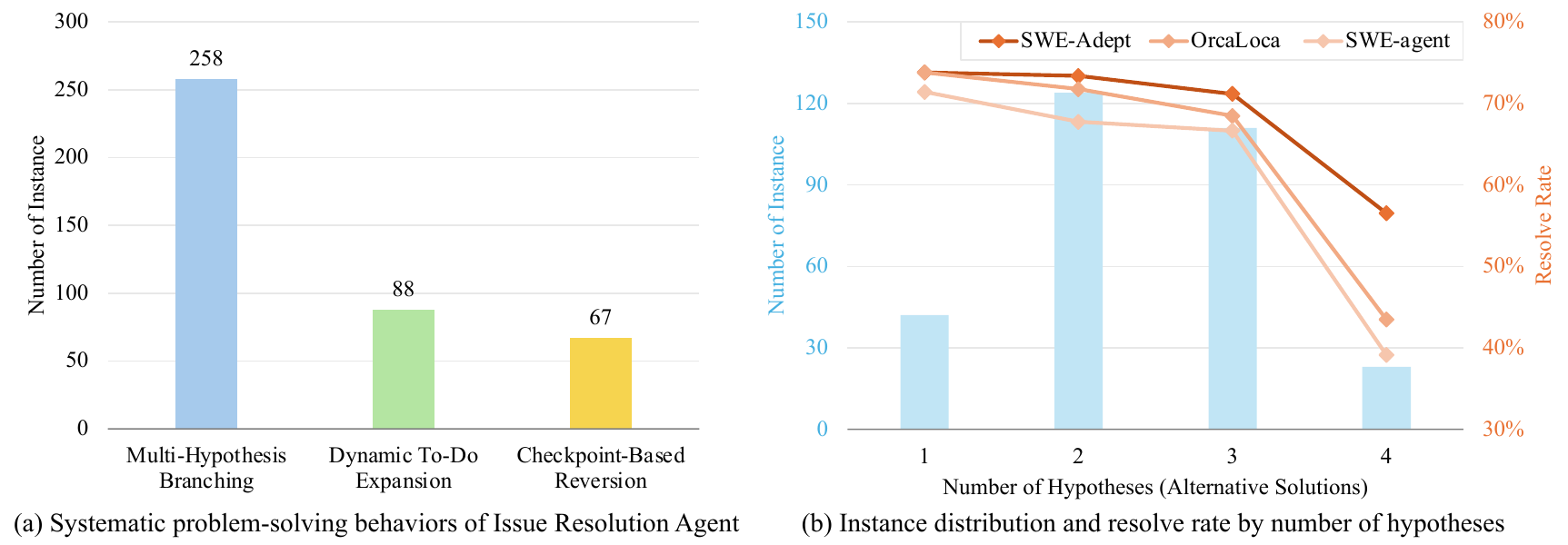}
\caption{Problem-solving behavior of \textbf{Issue Resolution Agent} and resolve rate by number of explored hypotheses. (a) Prevalence of systematic problem-solving behaviors: multi-hypothesis branching, dynamic to-do expansion, and checkpoint-based reversion (behaviors are non-mutually exclusive and may overlap). (b) Instance distribution (bars) and resolve rate (lines) by number of explored hypotheses on SWE-Bench Lite with Claude-Sonnet-4.5.}
\label{resolution_behavior}
\end{figure*}

\begingroup

\begin{table*}[t]
\small
\centering
\setlength{\tabcolsep}{4pt}
\renewcommand{\arraystretch}{1.2}

\begin{tabularx}{\textwidth}{>{\raggedright\arraybackslash}X c c}
\toprule
Framework & SWE-Bench Lite & SWE-Bench Pro \\
\midrule

SWE-agent
& 66.0\% & 39.5\% \\

\rowcolor[HTML]{f2f2f2}
SWE-Adept
& \textbf{71.3\%} & \textbf{45.0\%} \\

\quad - w/o designed issue localization
& 69.0\% & 43.3\% \\

\quad - w/o designed issue resolution
& 68.7\% & 41.5\% \\

\bottomrule
\end{tabularx}

\caption{Ablation study evaluating the contribution of each agent in SWE-Adept. Results report end-to-end resolve rate on SWE-Bench Lite and SWE-Bench Pro with Claude-Sonnet-4.5. ``w/o designed'' replaces the corresponding SWE-Adept module with the SWE-agent module.}
\label{resolve_rate_ablation}
\end{table*}

\endgroup
\begingroup

\begin{table*}[t]
\small
\centering
\setlength{\tabcolsep}{4pt}
\renewcommand{\arraystretch}{1.2}

\begin{tabularx}{\textwidth}{>{\raggedright\arraybackslash}X c c c c}
\toprule
& \multicolumn{2}{c}{SWE-Bench Lite}
& \multicolumn{2}{c}{SWE-Bench Pro} \\
\cmidrule(lr){2-3}\cmidrule(lr){4-5}
& Func Acc@5 & \# Tokens & Func Acc@5 & \# Tokens \\
\midrule

\multicolumn{5}{l}{\textbullet\ \textit{GPT-5.2}} \\

Search returns full code content
& 69.3\% & 218k & 45.0\% & 316k \\

\rowcolor[HTML]{f2f2f2}
Search returns code preview + specialized filtering
& \textbf{70.8\%} & 197k & \textbf{46.0\%} & 252k \\

\midrule

\multicolumn{5}{l}{\textbullet\ \textit{Claude-Sonnet-4.5}} \\

Search returns full code content
& 87.2\% & 393k & 62.0\% & 502k \\

\rowcolor[HTML]{f2f2f2}
Search returns code preview + specialized filtering
& \textbf{87.8\%} & 348k & \textbf{63.3\%} & 427k \\

\bottomrule
\end{tabularx}

\caption{Ablation study evaluating context management for \underline{issue localization}. ``Search returns full code content'' provides complete source code during traversal and lets the agent directly select candidate locations from the full-code context. ``Search returns code preview + specialized filtering'' represents our proposed design: it provides compact code previews during traversal, first filters preview-level candidate locations, and then loads full source code only for shortlisted locations during final re-ranking.}
\label{search_ablation}
\end{table*}

\endgroup
\begingroup

\begin{table*}[t]
\small
\centering
\setlength{\tabcolsep}{4pt}
\renewcommand{\arraystretch}{1.2}

\begin{tabularx}{\textwidth}{>{\raggedright\arraybackslash}X c c}
\toprule
Framework & SWE-Bench Lite & SWE-Bench Pro \\
\midrule

SWE-agent
& 66.0\% & 39.5\% \\

SWE-agent + raw Git commands (prompted)
& 67.0\% & 37.0\% \\

\midrule

\rowcolor[HTML]{f2f2f2}
SWE-Adept
& \textbf{71.3\%} & \textbf{45.0\%} \\

SWE-Adept + raw Git commands (no \texttt{hypothesis\_git}, no memory)
& 69.0\% & 41.5\% \\

\bottomrule
\end{tabularx}

\caption{Ablation study evaluating raw Git command usage for \underline{issue resolution}. Results report end-to-end resolve rate on SWE-Bench Lite and SWE-Bench Pro with Claude-Sonnet-4.5.}
\label{raw_git_ablation}
\end{table*}

\endgroup

\begingroup

\begin{table*}[t]
\small
\centering
\setlength{\tabcolsep}{4pt}
\renewcommand{\arraystretch}{1.2}

\begin{tabularx}{\textwidth}{>{\raggedright\arraybackslash}X c c c c}
\toprule
\multirow{2.5}{*}{Framework}
  & \multicolumn{2}{c}{SWE-Bench Lite}
  & \multicolumn{2}{c}{SWE-Bench Pro} \\
\cmidrule(lr){2-3}\cmidrule(lr){4-5}
  & \textit{Localize} & \textit{Resolve} & \textit{Localize} & \textit{Resolve} \\
\midrule

mini-SWE-Agent
& 75.9\% & 64.2\% & 50.5\% & 36.8\% \\

Claude Code
& 85.4\% & 70.2\% & 59.5\% & 42.5\% \\

\midrule

\rowcolor[HTML]{f2f2f2}
\textbf{SWE-Adept}
& \textbf{87.8\%} & \textbf{71.3\%}
& \textbf{63.3\%} & \textbf{45.0\%} \\

\bottomrule
\end{tabularx}

\caption{Comparison with Claude Code and mini-SWE-Agent using Claude-Sonnet-4.5. \textit{Localize} denotes function-level localization accuracy (Func Acc@5), and \textit{Resolve} denotes end-to-end issue resolve rate.}
\label{compare_with_cc}
\end{table*}

\endgroup

\begingroup

\begin{table*}[t]
\small
\centering
\setlength{\tabcolsep}{4pt}
\renewcommand{\arraystretch}{1.2}

\begin{tabularx}{\textwidth}{>{\raggedright\arraybackslash}X >{\raggedright\arraybackslash}X c c c c}
\toprule
\multirow{2.5}{*}{Framework} 
  & \multirow{2.5}{*}{Model} 
  & \multicolumn{2}{c}{SWE-Bench Lite} 
  & \multicolumn{2}{c}{SWE-Bench Pro} \\
\cmidrule(lr){3-4}\cmidrule(lr){5-6}
  & & \textit{Localize} & \textit{Resolve} & \textit{Localize} & \textit{Resolve} \\
\midrule

SWE-agent
& GLM-4.7-Flash (30B)
& 55.5\% & 43.3\% & 35.0\% & 26.5\% \\

SWE-agent
& Qwen3.5-27B
& 67.3\% & 56.8\% & 43.5\% & 33.8\% \\

\midrule

\rowcolor[HTML]{f2f2f2}
\textbf{SWE-Adept}
& GLM-4.7-Flash (30B)
& \textbf{63.3\%} & \textbf{48.2\%}
& \textbf{42.0\%} & \textbf{30.3\%} \\

\rowcolor[HTML]{f2f2f2}
\textbf{SWE-Adept}
& Qwen3.5-27B
& \textbf{75.4\%} & \textbf{62.7\%}
& \textbf{51.0\%} & \textbf{38.3\%} \\

\bottomrule
\end{tabularx}

\caption{Issue localization and end-to-end issue resolution results on SWE-Bench Lite and SWE-Bench Pro using open-source models. \textit{Localize} denotes function-level localization accuracy (Func Acc@5), and \textit{Resolve} denotes end-to-end issue resolve rate.}
\label{open_source_models}
\end{table*}

\endgroup

\begingroup

\newcolumntype{P}[1]{>{\raggedright\arraybackslash}p{#1}}

\begin{table*}[t]
\small
\centering
\setlength{\tabcolsep}{4pt}
\renewcommand{\arraystretch}{1.1}
\begin{tabularx}{\textwidth}{l X P{0.22\textwidth} P{0.30\textwidth}}
\toprule
\textbf{Tool} & \textbf{Description} & \textbf{Input Parameters} & \textbf{Output (Agent Observation)} \\
\midrule

\texttt{find\_file} &
Search \underline{file} by exact file-name or \texttt{glob} pattern. 
\vspace{2pt}

\color{black!35} Enumerate files within codebase (optionally
restricted to a specified directory).
&
\begin{tabular}[t]{@{}l@{}}
\textit{file\_name}\textsuperscript{\dag} \\
\textit{dir\_path}\textsuperscript{\ddag}
\end{tabular} &
Matched file paths; for each file, a skeleton listing the signatures of classes and functions defined in that file. \\
\addlinespace[3pt]\cdashline{1-4}\addlinespace[3pt]

\texttt{find\_code\_def} &
Search \underline{class/function} definition via exact matching, with fallback to \texttt{regex} matching and fuzzy matching ranked by a weighted sum of character-level similarity metrics (n-gram, Jaro–Winkler distance, and longest common subsequence). 
\vspace{2pt}

\color{black!35} Enumerate indexed code units (optionally restricted to a file).
&
\begin{tabular}[t]{@{}l@{}}
\textit{definition\_name}\textsuperscript{\dag} \\
\textit{file\_path}\textsuperscript{\ddag}
\end{tabular} &
Ranked retrieved code definitions with file path, line span, child-unit identifiers, and a concise code preview (definition signature + child-unit invocation context). \\
\addlinespace[3pt]\cdashline{1-4}\addlinespace[3pt]

\texttt{find\_code\_content} &
Search \underline{variable} name (robust to camel-case/snake-case variants) or an exact \underline{code snippet}. 
\vspace{2pt}

\color{black!35} Enumerate code lines (optionally restricted to a file or a line span).
&
\begin{tabular}[t]{@{}l@{}}
\textit{content}\textsuperscript{\dag} \\
\textit{file\_path}\textsuperscript{\ddag} \\
\textit{start\_line}\textsuperscript{\ddag} \\
\textit{end\_line}\textsuperscript{\ddag}
\end{tabular} &
Retrieved matches, each associated to a containing code unit (function/class/code chunk), with the unit name, file path, line span, child-unit identifiers, and a concise code preview (unit signature + matched lines). \\
\addlinespace[3pt]\cdashline{1-4}\addlinespace[3pt]

\texttt{find\_child\_unit} &
Search \underline{class/function} definition via exact matching given its name and file path (specified by a child-unit identifier). 
\vspace{2pt}

\color{black!35} Enumerate indexed code units restricted to a file.
&
\begin{tabular}[t]{@{}l@{}}
\textit{definition\_name}\textsuperscript{\dag} \\
\textit{file\_path}\textsuperscript{\dag}
\end{tabular} &
Exact-match class/function definition with file path, line span, child-unit identifiers, and a concise code preview (definition signature + child-unit invocation context). \\
\addlinespace[3pt]\cdashline{1-4}\addlinespace[3pt]

\texttt{finish\_search} &
Signal completion of the search and trigger subsequent filtering/ranking over the collected candidates. &
None &
None \\

\bottomrule
\end{tabularx}
\caption{List of tools for \textbf{Issue Localization Agent}. Superscripts of input parameters denote argument requirement: required\textsuperscript{\dag}, optional\textsuperscript{\ddag}. Gray text in the Description column provides auxiliary information on how each search tool performs enumeration.}
\label{localization_tool}
\end{table*}

\endgroup
\begingroup

\begin{figure*}[t]
\centering
\begin{tcolorbox}[
  width=\textwidth,
  title={Prompts for Issue Localization Agent},
  colback=gray!5,
  colframe=nmgray!75!black,
  fontupper=\footnotesize\linespread{1.1}\selectfont
]
\noindent
\textbf{System Prompt:}

\vspace{3pt}
You are a specialized issue localization agent responsible for conducting systematic search across code repositories to identify the code locations related to a given GitHub issue.
      
\vspace{3pt}
\#\# Follow this systematic Localization Workflow

\vspace{3pt}
\#\#\# Phase 1: Issue Analysis \& Entry Point Identification
\vspace{3pt}

-- Issue classification: Classify the issue as a bug fix, feature addition, performance issue, or configuration problem to set the localization focus. Use this category to prioritize likely modules, files, and configuration points.

-- Entry point extraction: Identify a shortlist of keywords to start searching, including any files/classes/functions named in the issue, and any locations suggested by error messages or stack traces.

-- Search plan: Specify an initial exploration order over entry points. Keep the plan adaptive by adding newly discovered entry points and dropping issue-irrelevant paths as results accumulate.

\vspace{3pt}
\#\#\# Phase 2: Agentic Depth-First Traversal
\vspace{3pt}
      
-- Locate the entry point: Use \texttt{find\_file}, \texttt{find\_code\_def}, or \texttt{find\_code\_content} to locate the current entry point. Use the returned file skeleton/code preview and child-unit identifiers to understand the local structure and how child units are invoked.

-- Selective deep search: Inspect child units and select only those that are likely related to the issue based on their names and invocation context. Explore following a selected branch via \texttt{find\_child\_unit} and repeat recursively, going deeper only when it helps localization; stop a branch when it becomes clearly unrelated or sufficiently understood.

-- Move to the next entry point: After exploring the relevant branches of one entry point, continue with the next entry point and repeat the previous steps.

-- End the search phase: Call \texttt{finish\_search} when the searched locations are sufficient.
      
\vspace{3pt}      
\#\#\# Phase 3: Result Evaluation \& Filtering
\vspace{3pt}

**First-stage Filter and Rank (Code-Preview and Location Heuristics)**
      
Use the available signals (file paths, line spans, definition names, child-unit lists, and code skeleton/previews) to prune and rank candidate locations. Prioritize locations that are directly mentioned or strongly implied by the issue, then keep nearby supporting code that is likely involved based on name/path semantics and invocation context; discard candidates with weak or unrelated signals. Output a high-to-low ranking.

\vspace{3pt}
**Second-stage Refine and Re-Rank (Content-Based Analysis)**

After Stage 1, the system will provide the full source code for the selected locations for further inspection. Confirm which locations actually result in the reported behavior. Refine the shortlist: drop locations that are clearly unrelated by implementation, and keep those with reasonable relevance. Output a final high-to-low ranked list.

\vspace{3pt}
\textbf{Instance Template:}
\vspace{3pt}

\texttt{\{\{repo\_name\}\}} \quad \graynote{\# Replaced with the repository name}

\vspace{3pt}
\texttt{\{\{issue\_description\}\}} \quad \graynote{\# Replaced with the issue description}

\vspace{3pt}
\textbf{Next Action Template:}
\vspace{3pt}

\texttt{\{\{observation\}\}} \quad \graynote{\# Replaced with the latest tool output}

\vspace{3pt}
Based on this observation, decide your next action.

\vspace{3pt} 
\textbf{First-Stage Filter Template:}
\vspace{3pt}

Based on the search results above, please perform the first-stage filtering and ranking using code-preview and location heuristics. Provide your results with locations ranked from highest to lowest relevance priority.
    
\vspace{3pt} 
\textbf{Second-Stage Filter Template:}
\vspace{3pt}

Here is the source code content for the locations you identified in  the first stage:

\vspace{3pt}
\texttt{\{\{source\_code\_content\}\}} \quad \graynote{\# Replaced with source code content}

\vspace{3pt}
Now perform the second-stage filtering and re-ranking using content-based analysis. Based on your analysis of actual code structure and logic, filter and re-rank the final locations from highest to lowest relevance.

\end{tcolorbox}
\caption{System prompt and stepwise instruction template for \textbf{Issue Localization Agent}.}
\label{localization_agent_prompt}
\end{figure*}

\endgroup

\begingroup

\newcolumntype{P}[1]{>{\raggedright\arraybackslash}p{#1}}
\newcolumntype{C}[1]{>{\centering\arraybackslash}p{#1}}

\newcommand{\CLI}[1]{\quad -{}- \textit{#1}}
\newcommand{\GITWRAPS}[1]{%
\\[1.0ex]%
{\color{black!35}\footnotesize%
\begin{tabular}[t]{@{}l@{}}
\textit{(internally executes)}\\
#1
\end{tabular}}%
}

\begin{table*}[t]
\small
\centering
\setlength{\tabcolsep}{4pt}
\renewcommand{\arraystretch}{1.10}

\begin{tabularx}{\textwidth}{P{0.17\textwidth} P{0.23\textwidth} X P{0.235\textwidth}}
\toprule
\textbf{Tool Family} & \textbf{Command (CLI-Based)} & \textbf{Description} & \textbf{Output (Agent Observation)} \\
\midrule

\multirow{3}{*}[-25ex]{\parbox{0.17\textwidth}{\texttt{hypothesis\_plan}}} &
\begin{tabular}[t]{@{}l@{}}
\texttt{update\_hypothesis}\\
\CLI{hypotheses\_markdown}
\end{tabular}
&
\textbullet\ Create, progress-track, and update hypotheses (i.e., alternative solutions) sorted by agent-estimated confidence, and annotate with status tags:

\begin{tabular}[t]{@{}ll@{}}
\texttt{[ ]}~(pending)      & \texttt{[-]}~(in-progress)\\
\texttt{[v]}~(successful)   & \texttt{[!]}~(failed)
\end{tabular}

Support dynamic expansion for adaptive planning.

\vspace{2pt}
\textbullet\ Store each hypothesis's \texttt{content} and \texttt{status} in working memory.
&
Hypothesis overview with current status and ordering. \\
\addlinespace[3pt]\cdashline{2-4}\addlinespace[3pt]

&
\begin{tabular}[t]{@{}l@{}}
\texttt{update\_todo}\\
\CLI{current\_hypothesis}\\
\CLI{todos\_markdown}
\end{tabular}
&
\textbullet\ Create, progress-track and update a to-do list for the current hypothesis, and annotate with status tags:

\begin{tabular}[t]{@{}ll@{}}
\texttt{[ ]}~(pending)      & \texttt{[-]}~(in-progress)\\
\texttt{[v]}~(successful)   & \texttt{[!]}~(failed)
\end{tabular}

Support dynamic expansion for adaptive planning.

\vspace{2pt}
\textbullet\ Store each to-do's \texttt{content} and \texttt{status} in working memory.
&
Hypothesis-corresponded to-do list with current status. 
\\
\addlinespace[3pt]\cdashline{2-4}\addlinespace[3pt]

&
\begin{tabular}[t]{@{}l@{}}
\texttt{log\_insight}\\
\CLI{insight}
\end{tabular}
&
\textbullet\ Generate insights for the current hypothesis based on execution feedback, which inform reflection-driven actions (e.g., revert) and final cross-hypothesis comparison.

\vspace{2pt}
\textbullet\ Store the insights in working memory.
&
Insights attached to the current hypothesis.
\\

\bottomrule
\end{tabularx}

\caption{Planning tools for \textbf{Issue Resolution Agent}. The \texttt{hypothesis\_plan} tool family interfaces with backend working memory to support adaptive planning, progress tracking, and insight logging. Tool invocations follow the command-line interface (CLI) format: each command uses -{}-\textit{parameter <value>} syntax, e.g., \texttt{hypothesis\_plan log\_insight} -{}-\textit{insight <insight\_content>}.}
\label{resolution_tool_1}
\end{table*}

\begin{table*}[t]
\small
\centering
\setlength{\tabcolsep}{4pt}
\renewcommand{\arraystretch}{1.10}

\begin{tabularx}{\textwidth}{P{0.17\textwidth} P{0.23\textwidth} X P{0.235\textwidth}}
\toprule
\textbf{Tool Family} & \textbf{Command (CLI-Based)} & \textbf{Description} & \textbf{Output (Agent Observation)} \\
\midrule

\multirow{6}{*}[-50ex]{\parbox{0.17\textwidth}{\texttt{hypothesis\_git}}} &
\begin{tabular}[t]{@{}l@{}}
\texttt{init\_base}
\GITWRAPS{
\texttt{git config}\\
\texttt{git rev-parse HEAD}\\
\texttt{git add -A}\\
\texttt{git commit -m}\\
\texttt{git rev-parse HEAD}
}
\end{tabular}
&
\textbullet\ Obtain the Git hashes of (i) the original code state and (ii) a shared common working base \texttt{git\_hash\_base} (including issue reproduction code) for subsequent hypothesis branches.

\vspace{2pt}
\textbullet\ Store the Git hashes in working memory.
&
Confirmation of saved original code state and common working base. \\
\addlinespace[3pt]\cdashline{2-4}\addlinespace[3pt]

&
\begin{tabular}[t]{@{}l@{}}
\texttt{start\_hypothesis}\\
\CLI{branch\_name}
\GITWRAPS{
\texttt{git stash push -m}\\
\texttt{git checkout}\\
\texttt{git checkout -b}
}
\end{tabular}
&
\textbullet\ Checkout a new hypothesis branch from the common working base.

\vspace{2pt}
\textbullet\ Retrieve the Git hash of the common working base from working memory.

\vspace{2pt}
\textbullet\ Store the branch name for the corresponding hypothesis in working memory.
&
Confirmation of branch creation and checkout; workspace moves to the hypothesis branch. \\
\addlinespace[3pt]\cdashline{2-4}\addlinespace[3pt]

&
\begin{tabular}[t]{@{}l@{}}
\texttt{commit\_todo}\\
\CLI{todo\_content}\\
\CLI{commit\_message}
\GITWRAPS{
\texttt{git add -A}\\
\texttt{git commit -m}\\
\texttt{git rev-parse HEAD}
}
\end{tabular}
&
\textbullet\ Commit code changes for one completed to-do step and obtain its checkpoint metadata (Git hash and commit message).

\vspace{2pt}
\textbullet\ Store the Git hash and commit message for the corresponding to-do in working memory.
&
Confirmation of to-do commit. \\
\addlinespace[3pt]\cdashline{2-4}\addlinespace[3pt]

&
\begin{tabular}[t]{@{}l@{}}
\texttt{revert\_to}\\
\CLI{source\_hypothesis}\\
\CLI{source\_todo}\\
\CLI{new\_branch\_name}
\GITWRAPS{
\texttt{git stash push -m}\\
\texttt{git checkout}\\
\texttt{git checkout -b}
}
\end{tabular}
&
\textbullet\ Checkout the specified to-do checkpoint and create a new branch from that state, then switch the workspace to the new branch and continue exploration under the new hypothesis branch.

\vspace{2pt}
\textbullet\ Retrieve the Git hash of the specified to-do from working memory.

\vspace{2pt}
\textbullet\ Store the new hypothesis's branch name in working memory.
&
Confirmation of revert and new-branch creation. \\
\addlinespace[3pt]\cdashline{2-4}\addlinespace[3pt]

&
\begin{tabular}[t]{@{}l@{}}
\texttt{compare\_hypotheses}
\GITWRAPS{
\texttt{git diff -{}-shortstat}\\
\texttt{git diff -{}-numstat}
}
\end{tabular}
&
Retrieve records from working memory and compare implemented hypotheses.
&
Hypothesis comparison report
\begin{tabular}[t]{@{}l@{}}
- branch name\\
- hypothesis content and status\\
- to-dos content and status\\
- commit messages\\
- insights\\
- code diff statistics\\
\hspace{1ex}(against the original state)
\end{tabular}
\\
\addlinespace[3pt]\cdashline{2-4}\addlinespace[3pt]

&
\begin{tabular}[t]{@{}l@{}}
\texttt{merge\_solution}\\
\CLI{branch\_name}
\GITWRAPS{
\texttt{git checkout -{}-detach}\\
\texttt{git merge}
}
\end{tabular}
&
\textbullet\ Retrieve the selected hypothesis branch and the Git hash of the original code state from working memory.

\vspace{2pt}
\textbullet\ Check out the original code state, then apply the selected branch's changes to produce a clean, patch-ready state for submission.
&
Confirmation that changes are applied onto the original code state. \\
\bottomrule
\end{tabularx}

\caption{Version-control tools for \textbf{Issue Resolution Agent}. The \texttt{hypothesis\_git} tool family interfaces with backend working memory to manage code-state information. Tool invocations follow the command-line interface (CLI) format: each command uses -{}-\textit{parameter <value>} syntax, e.g., \texttt{hypothesis\_git merge\_solution} -{}-\textit{branch\_name <name>}. Gray text in the Command column lists the raw Git commands executed internally (not exposed to the agent).}
\label{resolution_tool_2}
\end{table*}

\endgroup
\begingroup
\begin{figure*}[t]
\centering
\begin{tcolorbox}[
  width=\textwidth,
  title={Prompts for Issue Resolution Agent},
  colback=gray!5,
  colframe=nmgray!75!black,
  fontupper=\footnotesize\linespread{1.1}\selectfont
]
\noindent
\textbf{System Prompt:}

\vspace{3pt}
You are a specialized issue resolution agent that can interact with a computer to solve repository-level software engineering tasks. A separate issue localization agent has already performed systematic codebase search and provided ranked code-location hints. Your job is to analyze the root cause leveraging the hinted locations, reproduce the issue, and implement a robust fix.

\vspace{3pt}
\#\# Follow this systematic Resolution Workflow

\vspace{3pt}
\#\#\# Phase 1: Analysis \& Planning
\vspace{3pt}

-- Location-hint analysis and problem reproduction: Start from the code locations provided and understand how they relate to the described issue. Create a script to reproduce the issue and execute it using the bash tool to confirm the error.

-- Deep investigation and root cause identification: Use the reproduction result to trace the failing execution flow, leverage hinted code locations to identify where the issue originates.

-- Multi-hypothesis planning: Decide adaptively between a single hypothesis (only when the root cause and fix strategy are clear) and multiple hypotheses (2-4) when the fix location/strategy is uncertain, spans multiple files/functions, or involves complex code hierarchies or test implications. Create hypotheses using \texttt{hypothesis\_plan update\_hypothesis}. For each hypothesis, specify a brief hypothesis description and a confidence score (0.1-1.0). 

\vspace{3pt}
\#\#\# Phase 2: Git-Based To-Do Implementation
\vspace{3pt}

-- Git workflow initialization: Invoke \texttt{hypothesis\_git init\_base} to store the original code state and create a common working base (includes reproduction scripts). Regular hypothesis branches start from this common working base; hypotheses created after \texttt{revert\_to} action start from specified to-do's code-state checkpoints.

-- Hypothesis branching and to-dos initialization: Create a descriptive branch name and switch the workspace to this branch. Mark the current hypothesis as \texttt{[-]} (in-progress). Then plan 2-4 initial to-dos via \texttt{hypothesis\_plan update\_todo}. Each to-do item must be either edit or test action. These initial to-dos reflect your best current plan.

-- To-do-based implementation with checkpointing: Execute one to-do at a time for the current hypothesis: (1) mark the current to-do as \texttt{[-]} (in-progress), (2) perform the action (edit or test), (3) checkpoint with \texttt{hypothesis\_git commit\_todo} (provide descriptive information, e.g., test results, in the commit message), and (4) mark the current to-do as \texttt{[v]} (successful) or \texttt{[!]} (failed). Enforce “one to-do = one commit”.

-- Dynamic to-do expansion when needed: Expand the to-do list for the active hypothesis only when implementation feedback indicates missing steps or uncovered edge cases without contradicting the hypothesis's core fix strategy.

-- Hypothesis exploration and completion: After finishing all to-dos for a hypothesis, validate thoroughly and mark it as [v] (successful) or [!] (failed). Log actionable insights whenever they arise. If insights suggest a partially correct hypothesis (i.e., earlier steps remain useful but later direction is wrong), revert to the appropriate prior to-do checkpoint (via \texttt{hypothesis\_git revert\_to}), create a new branch, and continue exploration under the new hypothesis branch. Repeat until all hypotheses have been implemented and evaluated—do not stop early even if one appears to work. If none succeeds, formulate new hypotheses and keep exploration.

\vspace{3pt}
\#\#\# Phase 3: Solution Finalization
\vspace{3pt}

-- Hypothesis comparison: Invoke \texttt{hypothesis\_git compare\_hypotheses} to review the aggregated history (status, to-dos, commits, insights, and code diffs) of hypotheses, then select the best solution.

-- Solution integration: Merge/apply the selected branch's changes to the original code state to produce a patch-ready state for submission.

\vspace{3pt}
\textbf{Instance Template:}
\vspace{3pt}

\texttt{\{\{repo\_name\}\}} \quad \graynote{\# Replaced with the repository name}

\vspace{3pt}
\texttt{\{\{issue\_description\}\}} \quad \graynote{\# Replaced with the issue description}

\vspace{3pt}
\texttt{\{\{code\_location\_hints\}\}} \quad \graynote{\# Replaced with the ranked locations}

\vspace{3pt}
\textbf{Next Action Template:}

\vspace{3pt}
\texttt{\{\{observation\}\}} \quad \graynote{\# Replaced with the latest tool output}

\vspace{3pt}
Based on this observation, decide your next action.

\vspace{3pt}
\textbf{Submission Template:}

\vspace{3pt}
Here is a list of all your changes:

\vspace{3pt}
\texttt{\{\{code\_diff\}\}} \quad \graynote{\# Replaced with the code diffs}

\vspace{3pt}
1. Remove your generated reproduction/test script.

2. If you have modified any original test files, restore them to the initial state.

3. Finally, run the \texttt{submit} command.

\end{tcolorbox}
\caption{System prompt and stepwise instruction template for \textbf{Issue Resolution Agent}.}
\label{resolution_agent_prompt}
\end{figure*}
\endgroup

\endgroup

\end{document}